\pdfoutput=1
\documentclass{Interspeech}

\interspeechcameraready

\title{Synthesizing speech with selected perceptual voice qualities - A case study with creaky voice}

\author[affiliation={1}]{Frederik}{Rautenberg}
\author[affiliation={2}]{Fritz}{Seebauer}
\author[affiliation={2}]{Jana}{Wiechmann}
\author[affiliation={1}]{Michael}{Kuhlmann}
\author[affiliation={2}]{Petra}{Wagner}
\author[affiliation={1}]{Reinhold}{Haeb-Umbach}

\affiliation{Department of Communications Engineering}{Paderborn University}{Germany}
\affiliation{Phonetics Work Group, Faculty of Linguistics and Literary Studies}{Bielefeld University}{Germany}
\email{\{rautenberg, kuhlmann, haeb\}@nt.uni-paderborn.de,\\\ \{fritz.seebauer, 
jana.wiechmann, 
petra.wagner\}@uni-bielefeld.de}

\usepackage{comment}

\usepackage[nosort]{cite}   
\usepackage[acronym]{glossaries}
\newacronym{cpp}{CPP}{Cepstral Peak Prominence}
\newacronym{mae}{MAE}{Mean Absolute Error}
\newacronym{hnr}{HNR}{Harmonic-to-Noise Ratio}
\newacronym{tts}{TTS}{Text-to-Speech}
\newacronym{ccnf}{CCNF}{Conditional Continuous Normalizing Flow}
\newacronym{elbo}{ELBO}{Evidence Lower Bound}
\newacronym{ode}{ODE}{Ordinary Differential Equation}
\newacronym{mos}{MOS}{Mean Opinion Score}
\newacronym{smos}{SMOS}{Speaker Similarity MOS}
\newacronym{pvq}{PVQ}{perceptual voice quality}
\usepackage{tikz}
\usepackage{stackengine}
\usetikzlibrary{shapes.geometric, calc, fit, positioning, patterns}
\newcommand{\vect}[1]{\ensuremath{\boldsymbol{\mathbf{#1}}}}
\usepackage{pgfplots}
\usepackage{cleveref}
\usepackage{multirow}
\usepackage{dirtytalk}
\usepgfplotslibrary{groupplots}

\newcommand{\meanstd}[2]{\num[round-mode=places, round-precision=1]{#1} {\footnotesize \textcolor{gray}{$\pm$\,\num[round-mode=places, round-precision=1]{#2}}}}
\newcommand{\meanstdtwo}[2]{\num[round-mode=places, round-precision=0]{#1} {\footnotesize \textcolor{gray}{$\pm$\,\num[round-mode=places, round-precision=0]{#2}}}}

\newcommand{\loga}{\mathrm{log}}

\begin{document}

\maketitle

\begin{abstract}
    
    The control of perceptual voice qualities in a text-to-speech (TTS) system is of interest for applications where unmanipulated and manipulated speech probes can serve to illustrate phonetic concepts that are otherwise difficult to grasp. Here, we show that a TTS system, that is augmented with a global speaker attribute manipulation block based on normalizing flows\footnote{Code available at \href{https://github.com/fgnt/pvq\_manipulation}{https://github.com/fgnt/pvq\_manipulation}}, is capable of correctly manipulating the non-persistent, localized quality of creaky voice, thus avoiding the necessity of a, typically unreliable, frame-wise creak predictor. Subjective listening tests confirm successful creak manipulation at a slightly reduced MOS score compared to the original recording.
\end{abstract}

\section{Introduction}
In the last few years, the quality of speech synthesis and voice conversion systems has reached a level of naturalness that is essentially on par with human speech \cite{tan2024naturalspeech}. Recent works that integrate prosody or emotional control even allow for the generation of specific speaking styles, such as spontaneous speech and personalized voices \cite{triantafyllopoulos2023overview}. 
In the study presented here, we focus on synthesizing speech with selected \glspl{pvq}, where \textit{creak} serves as a prototypical example. The modification of such specific voice characteristics and attributes has so far attracted comparatively low attention \cite{anastassiou2024voiceshop, netzorg2023permod, rautenberg2025speech, li2024articulatory}.

Our motivation for this work is to support expert phoneticians in training students on the perceptual and acoustic-phonetic properties of \glspl{pvq}. 
We aim to design a TTS system that can generate speech probes with predefined \glspl{pvq}, where the perceived speaker identity should not change when manipulating them. Further, we wish to have zero-shot capability, allowing us to modify the voice of a speaker not seen in training. 
We have chosen \textit{creak} for our study, because, unlike the \glspl{pvq} investigated in \cite{rautenberg2025speech}, it is typically non-persistent and often occurs locally within an utterance, e.g., utterance finally. Also, we can compare our results with existing prior research.

\textit{Creaky voice} is characterized by a low rate of vocal fold vibration, combined with a constricted glottis, resulting in a low and irregular pitch,  \cite{keating2015acoustic}. While the aforementioned pattern is defined as the prototypical form of \textit{creaky} voice, there also exist atypical variations of creak, with their distinct acoustic properties \cite{keating2015acoustic}, making \textit{creak} hard to grasp analytically. 
However, as \textit{creak} fulfills many communicative and sociolinguistic functions, its analysis is vital within speech science, but has so far received only little attention in the field of speech synthesis or voice editing. 
The authors of \cite{lameris2024role} modified a \gls{tts} system to control the presence of \textit{creak} in the speech signal. They introduced a conditioning mechanism based on word-level \textit{creak} percentages, allowing for manipulation of \textit{creak}. Their study examined two different types of \textit{creak} placement and analyzed its impact on social perception and turn-taking processes.  In \cite{lameris2023neural}, the code from \cite{lameris2023prosody} was adapted by replacing prosodic acoustic features with word-level \textit{creak} probabilities. They defined three distinct types of \textit{creak}: no-creak, stylistic creak, and end-of-phrase creak. Experimental results demonstrated that, given the conditioning, the model successfully synthesized speech that aligned with the specified \textit{creak} characteristics. 
In these works \cite{lameris2023neural, lameris2023prosody}, the \gls{tts} system was trained in a speaker-dependent manner.

In their most recent study \cite{lameris2024creakvc}, a pre-trained voice conversion system \cite{li2023freevc} in combination with a pre-trained WavLM model \cite{chen2022wavlm} was employed. The system was adapted and fine-tuned to enable \textit{creak} modification in synthesized speech.
Their approach utilizes frame-wise \textit{creak} probabilities as an additional conditioning factor. These probabilities are extracted using CreaPy \cite{paierl2023creapy}, a tool that analyzes acoustic features and employs a classifier to predict frame-wise \textit{creak} probabilities. By adjusting the conditioning, the level of \textit{creak} in specific regions of the speech can be controlled. 
However, the effectiveness of this method relies on accurate frame-wise \textit{creak} probability estimations, which are notoriously hard to obtain. Experiments in \cite{paierl2023creapy} showed that global \textit{creak} probability estimations demonstrate a significantly higher agreement with human annotations than frame-wise estimations, which motivates a manipulation technique that requires only global \textit{creak} probabilities. Recent works \cite{anastassiou2024voiceshop, rautenberg2025speech} have demonstrated the effectiveness of global speaker attribute modification in adjusting both speaker characteristics and \glspl{pvq}. These studies applied normalizing flows \cite{rezende2015variational}, allowing continuous control over global speaker attributes. 

The study presented here aims to investigate whether such global speaker attribute modification is appropriate for modifying non-persistent attributes like \textit{creak}, thus eliminating necessity of local \textit{creak} prediction.
Our investigations show that our model, which is based on \cite{rautenberg2025speech}, successfully places \textit{creak} modifications mainly in voiced segments by analyzing their influence on the \gls{tts} embedding representations. Indeed, as a phonation type, \textit{creaky voice} depends on the activity of the vocal folds and is therefore limited to voiced signal parts \cite{paierl2023creapy, laver1980phonetic},


While different types of \textit{creak} exist, we here focus on manipulating the prototypical form and for now do not take into account its conversational functions. 
This is in part caused by the fact that we use a corpus of read speech, LibriTTS-R \cite{koizumi2023libritts}, for training our system, which does not contain spontaneous speech or dialogues. We opted for this data set, because it is sufficiently large to allow synthesizing and manipulating the speech of speakers not seen in training, an important property for the application we are targeting.

We demonstrate that adjusting the global \textit{creak} probability effectively influences the perceived \textit{creakiness} in synthesized speech \footnote{Audio examples: \href{https://groups.uni-paderborn.de/nt/interspeech_2025_creak_demo/interspeech_creak_demo.html}{https://go.upb.de/Interspeech\_creak\_demo}}. To validate this, we conduct listening tests with phonetic experts. Given the limited research and open source models in this area, we compare our system to the most recent one \cite{li2023freevc}, which follows a different approach by relying on local \textit{creak} probabilities. Our results indicate that global manipulation achieves comparable outcomes to local manipulations, suggesting that a global approach is a possible alternative.


\section{Controlling voice quality in TTS} \label{sec:controlling}
We adapted a \gls{tts} system to modify \textit{creak}, a voice quality present in certain locations of speech, using a global speaker manipulation mechanism to apply modifications and ensure that the changes are correctly positioned.

\subsection{Adapting TTS for speaker control}
 Our approach is based on YourTTS \cite{casanova2022yourtts}, an extension of VITS \cite{kim2021conditional}. The model is trained to maximize the \gls{elbo}:
\begin{equation}
    \begin{aligned}
        \loga \, p_{\vect{X}}(\vect{x} | \vect{c}) \geq  \mathbb{E}_{q_{\vect{Z}}} \left[ \vphantom{\frac{test}{test}} \right.& \loga \,p_{\vect{X}}(\vect{x} | \vect{z}) - \left. \loga\frac{q_{\vect{Z}}(\vect{z}|\vect{x})}{p_{\vect{Z}}(\mathbf{z}|\vect{c})} \right] \, ,  
    \end{aligned}
\end{equation}
given the latent embedding $\vect{Z} \in \mathcal{R}^{D \times T} $, the conditioning $\vect{c} = \left[ \vect{c}_{\mathrm{text}}, \vect{s}\right]$, which is a combination of the text embedding $\vect{c}_{\mathrm{text}}$ and a speaker embedding $\vect{s}$, and the speech signal $\vect{x}$.
$p_{\vect{Z}}(\mathbf{z}|\vect{c})$ is the prior, $p_{\vect{X}}(\vect{x} | \vect{z})$ the likelihood and $q_{\vect{Z}}(\vect{z}|\vect{x})$ the posterior distribution. All distributions are approximated by parametrized models. The prior encoder consists of a text encoder followed by a projection layer, which estimates the representation of the text input. An alignment function maps the text representation to the estimated duration of the target speech. Furthermore, the encoder incorporates a normalizing flow that enhances the flexibility of the distribution \cite{kim2021conditional}. This flow consists of a stack of coupling layers and is designed such that the Jacobian determinant remains one, by applying a shift-only operation. 
A HiFi-GAN \cite{kong2020hifi} 
is used as the decoder, which synthesizes from the embedding $\vec{Z}$ the speech signal $\hat{\vec{x}}$. The input of the posterior encoder is the spectrogram of $\vect{x}$, the encoder is designed, such that the time resolution  of $\vect{Z}$ matches  that of the spectrogram. This encoder is only used during training \cite{kim2021conditional}. 

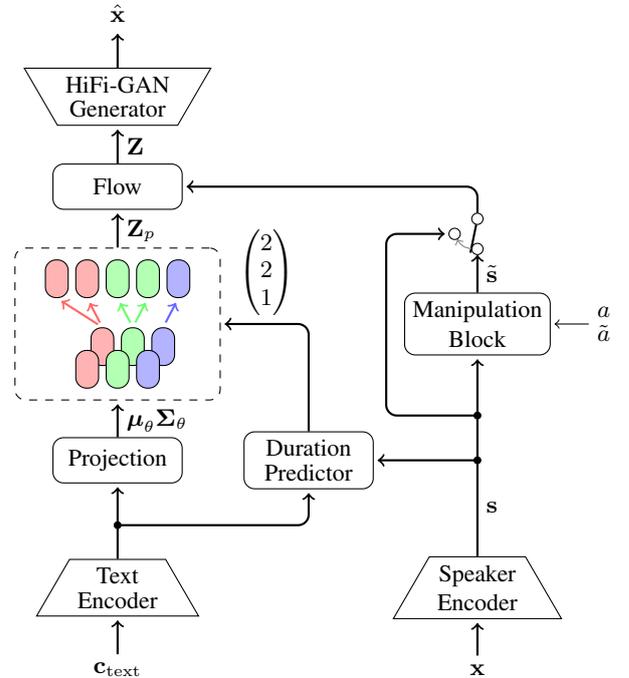
\begin{figure}[!b]
    \centering
    \vspace*{-0.5cm}
\tikzstyle{point} = [draw, fill=white, circle, radius=1cm, scale=0.5]
\tikzstyle{smallpoint} = [draw, fill=black, circle, radius=1cm, scale=0.3]
\tikzstyle{block} = [rectangle, draw, text centered, minimum width=1.7cm, minimum height=0.6cm, rounded corners]
\tikzstyle{smallblock} = [rectangle, draw, text centered, minimum width=0.3cm, minimum height=0.5cm, rounded corners, inner sep=2pt]
\tikzstyle{encoder} = [trapezium, draw,  text centered, minimum width=1.7cm, minimum height=0.6cm]

\def\groupOneYOffset{0.9}
\def\groupTwoYOffset{1.2}  
\def\groupThreeYOffset{2.1} 
\def\smallBlockSpacing{0.4}

\begin{tikzpicture}[auto, node distance=0.5cm, remember picture]
    \node [] (input_text) {$\vect{c}_{\mathrm{text}}$};
    \node[encoder, above = of input_text] (TE) {\stackanchor{Text}{Encoder}};

    \node[block, above = 1cm of TE] (AG) {Projection};

    \node[smallblock, fill=red!30] (G2A) at ($(AG.north) + (-\smallBlockSpacing+0.2, \groupTwoYOffset)$) {};
    \node[smallblock, fill=green!30] (G2B) at ($(AG.north) + (0+0.2, \groupTwoYOffset)$) {};
    \node[smallblock, fill=blue!30] (G2C) at ($(AG.north) + (\smallBlockSpacing+0.2, \groupTwoYOffset)$) {};
    \node[smallblock, fill=red!30] (G1A) at ($(AG.north) + (-\smallBlockSpacing, \groupOneYOffset)$) {};
    \node[smallblock, fill=green!30] (G1B) at ($(AG.north) + (0, \groupOneYOffset)$) {};
    \node[smallblock, fill=blue!30] (G1C) at ($(AG.north) + (\smallBlockSpacing, \groupOneYOffset)$) {};

    \node[smallblock, fill=red!30] (G3A) at ($(AG.north) + (-0.8, \groupThreeYOffset)$) {};
    \node[smallblock, fill=red!30] (G3B) at ($(AG.north) + (-0.4, \groupThreeYOffset)$) {};
    \node[smallblock, fill=green!30] (G3C) at ($(AG.north) + (0.0, \groupThreeYOffset)$) {};
    \node[smallblock, fill=green!30] (G3D) at ($(AG.north) + (0.4, \groupThreeYOffset)$) {};
    \node[smallblock, fill=blue!30] (G3E) at ($(AG.north) + (0.8, \groupThreeYOffset)$) {};

    \draw[->, thick, red!60, shorten >=2pt, shorten <=2pt, bend left=10] (G2A.north) -- (G3A.south);
    \draw[->, thick, red!60, shorten >=2pt, shorten <=2pt, bend right=10] (G2A.north) -- (G3B.south);
    
    \draw[->, thick, green!60, shorten >=2pt, shorten <=2pt, bend left=10] (G2B.north) -- (G3C.south);
    \draw[->, thick, green!60, shorten >=2pt, shorten <=2pt, bend right=10] (G2B.north) -- (G3D.south);
    
    \draw[->, thick, blue!60, shorten >=2pt, shorten <=2pt, bend left=10] (G2C.north) -- (G3E.south);

    \begin{scope}[remember picture]
        \node[fit=(G1B) (G3A) (G3E), draw, inner sep=0.15cm, rectangle, rounded corners, dashed, minimum width=2.7cm] (frame) {};
    \end{scope}
    \node[block, above = 0.5cm of frame] (FB) {Flow};
    \node[block, right = 0.8cm of AG] (DP) {\stackanchor{Duration}{Predictor}};
    \draw[->, thick, rounded corners] (DP.north) |- node[above, near end] {$\begin{pmatrix} 2 \\ 2 \\ 1 \end{pmatrix}$} ([shift=({0.05,0})] frame.east);
    \draw[->, thick] (AG.north) -- node[right] {$\vect{\mu}_\theta \vect{\Sigma}_\theta$}([shift=({0,-0.05})]frame.south);
    \node[encoder, shape border rotate = 180, above = of FB] (hifi) {\stackanchor{HiFi-GAN}{Generator}};
    \node[above=of hifi] (speech_signal_output) {$\hat{\vect{x}}$};

    \draw[->, thick] (input_text) -- ([shift=({0,-0.05})]TE.south);
    \draw[->, thick] (TE) -- ([shift=({0,-0.05})]AG.south);
    \draw[->, thick] (FB) -- node[right] {$\vect{Z}$} ([shift=({0,-0.05})]hifi.south);
    \draw[->, thick] (hifi.north) -- ([shift=({0,-0.05})]speech_signal_output.south);
    \draw[->, thick] (frame.north) -- node[right] {$\vect{Z}_p$} ([shift=({0,-0.05})]FB.south);

    \node[encoder, right = 3cm of TE] (SE) {\stackanchor{Speaker}{Encoder}};
    \node[block, rounded corners] at (SE.north |- frame.west) (ode) {\stackanchor{Manipulation}{Block}};
    \node at (input_text.east -| SE.south) (speech_signal){$\vect{x}$};

    \draw[->, thick] (speech_signal) -- ([shift=({0,-0.05})] SE.south);

    \node[smallpoint, above=0.4cm of TE] (sm2) {};
    \node[smallpoint] at (DP.east -| SE.north) (sm1) {};
    \draw[->, thick, rounded corners] (sm1) -- ([shift=({0.05, 0})] DP.east);
    \draw[->, thick] (SE.north) -- node[right, near start] {$\vect{s}$} ([shift=({0,-0.05})] ode.south);
    \draw[->, thick, rounded corners] (sm2) -| ([shift=({0,-0.05})]DP.south);
    \node[right = of ode](attribut){\stackanchor{$a$}{$\tilde{a}$}};
    \draw[->] (attribut.west) -- ([shift=({0.05, 0})]ode.east);

    \node[above = 0.5cm of ode.north, point] (point1) {};
    \node[point] at ($(point1) + (-0.3cm,0.2cm)$) (point2) {};
    \node[point] at ($(point1) + (0cm,0.4cm)$) (point3) {};
    \draw[-, thick] ([yshift=-0.05cm, xshift=-0.4] point1.west) -- (point3);
    \draw[->, color=gray] (point1) to [bend left] (point2);

    \draw[->, thick, rounded corners] (ode.north) -- node[right] {$\tilde{\vect{s}}$} (point1.south);

    \node[smallpoint, above = 0.5cm of sm1] (sm3) {};
    \draw[->, thick, rounded corners] (sm3)  -- ++(-1.2cm, 0.cm) |- ([shift=({-0.05, 0})] point2.west);
    \draw[->, thick, rounded corners] (point3) |- ([shift=({0.05, 0})] FB.east);
\end{tikzpicture}
    \vspace*{-0.5cm}
    \caption{\gls{tts} inference with a speaker embedding manipulation block, where $a$ is the creak probability of $\vect{x}$ and $a + \tilde{a}$ its modified probability. The switch controls whether the original or the modified speaker embedding is used.}
    \label{fig:tts}
\end{figure}

\Cref{fig:tts} illustrates the system during inference. Compared to YourTTS \cite{casanova2022yourtts}, we introduced several modifications to enhance control over speaker attributes. First, we removed the conditioning of the decoder on the speaker embedding $\vect{s}$. The motivation behind this change is to constrain the influence of the speaker embedding to a single fixed point in the model, ensuring a better investigation of its influence. Second, we replaced the original speaker encoder with a d-vector model \cite{Cord-Landwehr_Boeddeker_Zorilă_Doddipatla_Haeb-Umbach_2023}. Lastly, we included a speaker manipulation block, allowing for controlled modification of speaker attributes. It is important to note that the duration predictor remains constrained by the unmanipulated speaker embedding. This ensures that the speaking rate is unaffected by the manipulation.

\subsection{Modifying speaker representations}
The authors of \cite{anastassiou2024voiceshop} applied the concept of \gls{ccnf} \cite{abdal2021styleflow} to achieve a global speaker attribute manipulation. This concept was followed in \cite{rautenberg2025speech} to manipulate a global perceptual voice quality. We use this idea to manipulate a positional \gls{pvq}, i.e., a quality that is not persistent. Our approach consists of a global speaker manipulation, but we assume and later investigate, that the manipulation is done on the correct location in the speech signal. We also use the concept of \gls{ccnf} to apply a global speaker manipulation. The goal of the \gls{ccnf} is to learn a transformation  of the random variable such that the speaker embedding becomes normally distributed after applying it. This transformation function is learned from the data $\mathcal{S} = \left\{\vect{s}_n\right\}_{n=1}^N$ by maximizing the following log-likelihood function \cite{grathwohl2018ffjord}
\begin{align}
    &l = \sum_n \loga p_{\vect{S}} \left( \vect{s}_n | \vect{a}_n \right) \nonumber \\
    &= \sum_n \loga p_{\vect{Z}_0}\left( \vect{z}_n(t_0)\right) + \int_{t_1}^{t_0}\mathrm{tr}\left( \frac{\mathrm{d} f(\vect{z}(t), t, \vect{a}_n)}{\mathrm{d}\vect{z}(t)} \right) \mathrm{d}t
    \label{eq:likelihood_2}
\end{align}
with 
\begin{align}
    \label{eq:transformation}
    \vect{z}_n(t_0) = \vect{z}_n(t_1) + \int_{t_1}^{t_0}  f(\vect{z}(t), t, \vect{a}) \mathrm{d}t  \, .
\end{align}
With $\vect{z}_n(t_0) \sim \mathcal{N}(\mathbf{0}, \mathbf{I})$, the parametrized function $f(\cdot)$ and the initial condition is given by $\vect{s} = \vect{z}_n(t_1)$. Note, we assumed that the speaker embedding is indirectly conditioned by the speaker attribute, thus no additional input for the speaker encoder is needed. Computing the log-likelihood  requires solving two \gls{ode} problems, \Cref{eq:likelihood_2} and \Cref{eq:transformation}.

\begin{figure*}[!htb]
    \centering
    \newcommand{\plotwidth}{7.5cm}
    \newcommand{\plotheight}{4.cm}
    \input{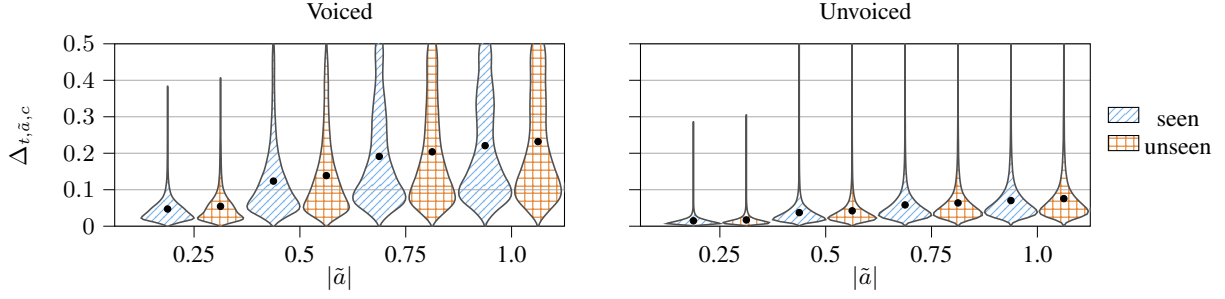}
    \vspace{-0.2cm}
    \caption{Distribution of differences $\Delta_{t,\tilde{a},c}$ across voiced and unvoiced classes for seen and unseen speakers. The plot combines positive and negative manipulations, with mean values highlighted to indicate overall trends.}
    \vspace{-0.2cm}
    \label{fig:diff_dist}
\end{figure*}

After training, the speaker embedding $\vect{s}$ is manipulated in the following steps. First, the speaker embedding $\vect{s}$ and its attribute vector $a$ are extracted from the speech signal $\vect{x}$. Next, the speaker embedding is transformed into $\vect{z}(t_0)$ by solving the \gls{ode} problem \Cref{eq:transformation} with the initial condition $\vect{s} = \vect{z}(t_1)$. Finally, the inverse of \Cref{eq:transformation} is applied, which is again an \gls{ode} problem. This time, the initial condition is set to the obtained vector $\vect{z}(t_0)$, but the attribute is modified to $a + \tilde{a}$, resulting in the manipulated speaker embedding $\tilde{\vec{s}}$. Together, these steps form the manipulation block, which enables controlled modification of speaker attributes.


\begin{table}[b]
\vspace*{-0.3cm}
  \caption{Mean and standard deviation of $\Delta_{t, \tilde{a},c} \cdot 10^2$ in voiced and unvoiced segments for different modification strengths with combined positive and negative manipulations}
  \label{tab:diff_z}
  \centering
  \begin{tabular}{ c  c  c c c c }
    \toprule
     \multirow{2}{*}{Set} & \multirow{2}{*}{Group} &\multicolumn{4}{c}{$|\tilde{a}|$}\\
    \cmidrule(l){3-6}
    &  & 0.25 & 0.5 & 0.75 & 1.0 \\
    \midrule
    \multirow{2}{*}{seen} & voiced & \meanstdtwo{5}{5} & \meanstdtwo{14}{11} & \meanstdtwo{20}{14} & \meanstdtwo{23}{15} \\
     & unvoiced & \meanstdtwo{2}{2}& \meanstdtwo{4}{4} &\meanstdtwo{6}{6} & \meanstdtwo{8}{6}\\
    \midrule
    \multirow{2}{*}{unseen} & voiced & \meanstdtwo{5}{4} & \meanstdtwo{12}{10} & \meanstdtwo{19}{14} & \meanstdtwo{22}{15}  \\
     & unvoiced & \meanstdtwo{1}{1} & \meanstdtwo{4}{3} & \meanstdtwo{6}{5} & \meanstdtwo{7}{6} \\
    \bottomrule
  \end{tabular}
\end{table}

\section{Experiments}
The experiments were conducted on the LibriTTS-R dataset \cite{koizumi2023libritts}, which applies sound quality improvements to the original LibriTTS dataset \cite{zen2019libritts}. LibriTTS-R comprises 585 hours of speech data from 2,456 speakers. The training of the \gls{tts} system followed the train-test split proposed in \cite{koizumi2023libritts}.

The manipulation block, consisting of a \gls{ccnf}, was trained independently of the \gls{tts} system using speaker embeddings $\vect{s}$ and their corresponding attribute $a$. The same train-test split as in the \gls{tts} system was used. Speaker conditioning denoted as $a$, represents the global \textit{creak} probability, which was extracted using CreaPy \cite{paierl2023creapy}. We followed the extraction process described in \cite{paierl2023creapy} but incorporated an additional energy-based voice activity detection  to preprocess the speech signal $\vect{x}$. This additional step was necessary to reduce the influence of noise in silent segments on the estimation. A frame-wise estimation of the \textit{creak} probability was then performed, with the final global estimation obtained by averaging the frame-wise results. The training objective was the likelihood, as explained in \Cref{eq:likelihood_2}. Optimizing this function required solving two \gls{ode} problems, for which we employed the solver from \cite{chen2018neural}. The trace estimation was performed using Hutchinson’s trace estimator \cite{grathwohl2018ffjord}. The function $f(\cdot)$ was modeled using a single CCNF block \cite{abdal2021styleflow} with a hidden size of 512. The \textit{creak} modification of the speech signal $\vect{x}$ was performed in the following steps: first, the speaker embedding $\vect{s}$ and its estimated attribute $a$ were extracted. Next, the manipulated speaker embedding $\tilde{\vect{s}}$ with the desired \textit{creak} probability was obtained by applying the manipulation block to $\vect{s}$. $\tilde{\vect{s}}$ was then used as input for the \gls{tts} system. The desired \textit{creak} measure was computed as the original estimated  $a$ plus a manipulation factor $\tilde{a}$. 

\subsection{Temporal analysis of creak manipulation}

Here, we are going to investigate whether the global speaker attribute manipulation is appropriate for manipulating the speech signal at the appropriate positions.
\textit{Creak} occurs only in voiced segments of speech \cite{paierl2023creapy, lameris-etal-2024-role}. 
To check, whether predominantly voiced segments are affected by the speaker attribute manipulation, we first extracted the text transcription of the unmanipulated speech signal $\hat{\vect{x}}$ using Whisper \cite{radford2023robust}. Then, we obtained phoneme annotations and their corresponding durations using the Montreal Forced Aligner \cite{mcauliffe17_interspeech}, employing a dictionary that extracts phonemes based on IPA charts. According to the IPA chart, phonetic experts categorized the phonemes into three groups: voiced, unvoiced, and silence, resulting in 42 voiced and 13 unvoiced phonemes. 

To determine which phonemes are most affected by \textit{creak} manipulation, we mapped the phonemes and their categories onto the time resolution of  $\mathbf{Z} = \begin{bmatrix} \vect{z}_1 & \dots & \vect{z}_T
\end{bmatrix}$ resulting in $\mathbf{Z}_c = \begin{bmatrix}
\vect{z}_{1,c} & \dots & \vect{z}_{T,c}
\end{bmatrix}$
where $c$ represents one of the three phoneme categories. The decoder is designed such that the synthesized spectrogram $\tilde{\vect{X}}$ maintains the same temporal resolution as the latent embedding $\vect{Z}$, using identical parameters as the posterior encoder during training. The same steps were applied to the manipulated speech signal $\tilde{\vect{x}}$. Using the \gls{mae} metric, we calculated the difference between unmanipulated and manipulated embeddings
\begin{align}
    \Delta_{t, \tilde{a}} = \frac{1}{D} \left\lVert\vect{z}_{t,a} - \vect{z}_{t,\tilde{a}} \right\rVert_1
\end{align}
with $\vect{z}_{t,a}$ the unmanipulated and $\vect{z}_{t,\tilde{a}}$ the manipulated embedding. Using the class label $c$ we categorized each difference to one of the three classes $\Delta_{t,\tilde{a},c}$. Note, we investigated the difference in the embedding space rather than in the synthesized voice, to ignore the effects of the decoder. 

\begin{table*}[!t]
\caption{\textit{Creak} across different strengths including the evaluation on the original speech. The table presents averaged Mean Opinion Score (MOS) ratings on a 5-point scale (1 = Bad, 5 = Excellent) and \textit{creak} ratings on a 100-point open ended interval scale (25 = no perceived creak, 75 = very strong creak)}
\vspace*{-0.5cm}
\begin{center}
\begin{tabular}{cccccc} 
\toprule
 \multirow{2}{*}{Method} & \multirow{2}{*}{Set} & \multirow{2}{*}{\stackanchor{Original}{Recording}} & \multicolumn{3}{c}{Creak Manipulation} \\
 \cmidrule(l){4-6} 
 & & & Suppressed & Unmanipulated & Amplified \\ 
 \midrule 
 \multirow{2}{*}{Proposed}& Perc. Creak (0-100) & \meanstd{44.08}{26.93} & \meanstd{25.38}{18.34} & \meanstd{39.12}{24.19} & \meanstd{74.19}{16.93} \\
 & MOS $\uparrow$ (1-5) & \meanstd{4.23}{0.97} & \meanstd{3.54}{1.13} & \meanstd{3.79}{1.24} & \meanstd{3.75}{1.25} \\
 \midrule 
 \multirow{2}{*}{CreakVC \cite{lameris2024creakvc}}& Perc. Creak (0-100) & \meanstd{39.35}{24.81} & \meanstd{25.23}{18.61} & \meanstd{42.23}{24.22} & \meanstd{85.25}{12.48} \\
 & MOS $\uparrow$ (1-5) & \meanstd{3.81}{0.96} & \meanstd{3.08}{1.16} & \meanstd{3.73}{1.05} & \meanstd{3.33}{1.12} \\
\bottomrule
\end{tabular}
\label{tab:results_perceived_pvq}
\end{center}
\vspace{-0.5cm}
\end{table*}

From 300 randomly selected utterances, both manipulated and unmanipulated embeddings were extracted, and the \gls{mae} was computed. \Cref{fig:diff_dist} visualizes the difference $\Delta_{t, \tilde{a}, c}$ as a function of the manipulation degree for utterances from the training and test set, while \Cref{tab:diff_z} presents the mean and standard deviation. As the manipulation factor increases, the magnitude of changes in the embeddings also grows, with distinct variations across phoneme categories. 
Voiced segments display much more noticeable differences than unvoiced segments, with both the mean difference and standard deviation increasing as the manipulation factor increases. 
While this does not prove that the manipulation corresponds to emphasizing or deemphasizing \textit{creak}, it indicates that the model's response to manipulation is particularly strong at the correct position, i.e., for voiced segments.

\begin{table}[!b]
  \caption{Pearson correlation coefficient $R$ between acoustic features extracted from $\tilde{\vect{x}}$ and $\tilde{a}$ for seen and unseen speakers}
  \label{tab:correlation_coef}
  \centering
  \begin{tabular}{c c c c c}
    \toprule
     Set & Creak & Pitch & HNR & H1-H2  \\
    \midrule
    seen &  0.81 & -0.80 & -0.90 & -0.59 \\
     unseen &  0.82 & -0.78 & -0.91 & -0.67 \\
    \bottomrule
  \end{tabular}
\end{table}

\subsection{Subjective listening tests}
We conducted a subjective listening test to evaluate the perceptual impact of \textit{creak} manipulation. We compared our system with CreakVC \cite{lameris2024creakvc}, which employs local \textit{creak} probabilities. Although both methods were fine-tuned on VCTK \cite{vctk2019}, our system showed reduced performance on that dataset. Therefore, we used speakers from LibriTTS-R for our proposed method and VCTK speakers for CreakVC.

As \textit{creak} is not a commonly known concept, we recruited 12 speech experts as participants. Given the time-intensive nature of the evaluation, each participant was presented with a randomized section of manipulated speech samples at three different modification levels: Suppressed, unmanipulated and amplified, corresponding to $\tilde{a} \in \left\{-1, 0, 1 \right\}$ for the proposed method and the mean average creak values $\tilde{a} \in \left\{-10, 0, 10 \right\}$ for CreakVC. These values were chosen to yield a similar degree of \textit{creak} in the synthesis measured with Creapy. For each manipulation, a speaker from the train set of each model is used. We made this choice, because CreakVC used nearly all speakers except one from the VCTK set for training. Each participant rated 16 samples from each system, resulting in 384 ratings in total, with an average audio duration of $5.48\,\mathrm{s}$ and covering 65 ($31\,$f, $34\,$m) unique speakers. Each speaker's original recording served as a reference, and presentation order of each trial was randomized.

The evaluation criteria comprised the perceived \textit{creak} and the \gls{mos} for perceived audio quality, assessed using the standard ITU-T scale \cite{ITU808} (1 = Bad to 5 = Excellent). For \textit{creak}, we employed a 100-point open-ended interval scale (25 = no creak, 75 = very strong creak), following the recommendations in \cite{kreiman2007and} and \cite{hinterleitner2011evaluation}.

\Cref{tab:results_perceived_pvq} presents the mean and standard deviation of the subjective ratings. Expert listeners clearly rated the amplified condition with higher creak levels and the suppressed condition with lower levels for both systems. Bonferonni corrected Wilcoxon rank sum tests \cite{bauer1972} confirmed that the differences in creak ratings were statistically significant $p<0.001$, with the exception of the unmanipulated to suppressed change (significant value at $p<0.05$ for the proposed method and $p<0.005$ for CreakVC). No significant difference was observed between the unmanipulated synthesis and the natural recording. Regarding the \gls{mos} ratings, a slight quality reduction was noted from the original to the synthesized unmanipulated speech, with CreakVC showing lower performance under both manipulation conditions. In summary, these findings indicate that our proposed global manipulation of \textit{creak} effectively modulates perceived \textit{creak}, achieving results comparable to those of a model that uses local \textit{creak} probabilities.

\subsection{Acoustic measurements}
Following \cite{keating2015acoustic}, a prototypical creaky voice is characterized by three acoustic properties: low pitch $f_0$, irregular pitch (measured by \gls{hnr}), and a constricted glottis (measured by the amplitude difference between the first and second harmonics (H1-H2)). Notably, three of the five features used by \cite{paierl2023creapy} to predict \textit{creak} probability correspond to these properties. In our objective test, we investigate whether our manipulated speech signals $\tilde{\vect{x}}$ differ in these acoustic measures. We synthesized voices with manipulation strengths $a \in \left\{-1.5, 0, 1.5 \right\}$ in increments of 0.25, and extracted the mean pitch (following \cite{morrison2023cross}), \gls{hnr}, and H1-H2 using Praat \footnote{www.praat.org}, and creak probability using Creapy \cite{paierl2023creapy}. \Cref{tab:correlation_coef} reports the Pearson correlation coefficient $R$ between $\tilde{a}$ and the corresponding acoustic features extracted from $\tilde{\vect{x}}$. A high positive correlation is observed between $\tilde{a}$ and mean \textit{creak} probability, while negative correlations are found for the other features, consistent with \cite{keating2015acoustic}. Although the correlations for pitch and \gls{hnr} are strong, the correlation for H1-H2 is less pronounced.

\section{Conclusions}
We could show that the system for global speaker attribute manipulation is able to manipulate the strength of \textit{creak}, although this voice quality is non-persistent and only locally present. Since the system does not employ any particular properties of \textit{creak}, we are confident that it can be used for the manipulation of a wide range of perceptual voice qualities with no or little adjustment.

\section{Acknowledgements}
\ifinterspeechfinal
Funded by the Deutsche Forschungsgemeinschaft (DFG, German Research Foundation): TRR 318/1 2021- 438445824 and 446378607.
\else
    Acknowledgements hidden during review.
\fi

\bibliographystyle{IEEEtran}
\bibliography{literature}

@inproceedings{lameris2024creakvc,
  title={{CreakVC: A Voice Conversion Tool for Modulating Creaky Voice}},
  author={Lameris, Harm and Gustafson, Joakim and Sz{\'e}kely, {\'E}va},
  booktitle={Interspeech 2024 Demo Session},
  year = {2024},
}

@inproceedings{li2023freevc,
  title={Freevc: Towards high-quality text-free one-shot voice conversion},
  author={Li, Jingyi and Tu, Weiping and Xiao, Li},
  booktitle={ICASSP 2023-2023 IEEE International Conference on Acoustics, Speech and Signal Processing (ICASSP)},
  pages={1--5},
  year={2023},
  organization={IEEE}
}

@inproceedings{paierl2023creapy,
  title={Creapy: A python-based tool for the detection of creak in conversational speech},
  author={Paierl, Michael and R{\"o}ck, Thomas and Wepner, Saskia and Kelterer, Anneliese and Schuppler, Barbara},
  booktitle={20th International Congress on Phonetic Sciences: ICPhS 2023},
  year={2023}
}

@inproceedings{lameris2024role,
  title={The Role of Creaky Voice in Turn Taking and the Perception of Speaker Stance: Experiments Using Controllable {TTS}},
  author={Lameris, Harm and Sz{\'e}kely, {\'E}va and Gustafson, Joakim},
  booktitle={Proceedings of the 2024 Joint International Conference on Computational Linguistics, Language Resources and Evaluation (LREC-COLING 2024)},
  pages={16058--16065},
  year={2024}
}

@inproceedings{keating2015acoustic,
  title={Acoustic properties of different kinds of creaky voice.},
  author={Keating, Patricia A and Garellek, Marc and Kreiman, Jody},
  booktitle={ICPhS},
  volume={1},
  pages={2--7},
  year={2015}
}

@inproceedings{netzorg2023permod,
  title={Permod: Perceptually grounded voice modification with latent diffusion models},
  author={Netzorg, Robin and Jalal, Ajil and McNulty, Luna and Anumanchipalli, Gopala Krishna},
  booktitle={2023 IEEE Automatic Speech Recognition and Understanding Workshop (ASRU)},
  pages={1--8},
  year={2023},
  organization={IEEE}
}

@article{laver1980phonetic,
  title={The phonetic description of voice quality},
  author={Laver, John},
  journal={Cambridge Studies in Linguistics London},
  volume={31},
  pages={1--186},
  year={1980}
}

@article{chen2022wavlm,
  title={Wavlm: Large-scale self-supervised pre-training for full stack speech processing},
  author={Chen, Sanyuan and Wang, Chengyi and Chen, Zhengyang and Wu, Yu and Liu, Shujie and Chen, Zhuo and Li, Jinyu and Kanda, Naoyuki and Yoshioka, Takuya and Xiao, Xiong and others},
  journal={IEEE Journal of Selected Topics in Signal Processing},
  volume={16},
  number={6},
  pages={1505--1518},
  year={2022},
  publisher={IEEE}
}

@inproceedings{mcauliffe17_interspeech,
  author={McAuliffe, Michael and Socolof, Michaela and Mihuc, Sarah and Wagner, Michael and Sonderegger, Morgan},
  title={{Montreal Forced Aligner: Trainable Text-Speech Alignment Using Kaldi}},
  year=2017,
  booktitle={Proc. Interspeech 2017},
  pages={498--502},
  doi={10.21437/Interspeech.2017-1386}
}

@inproceedings{koizumi2023libritts,
  title     = {{LibriTTS-R: A Restored Multi-Speaker Text-to-Speech Corpus}},
  author    = {Yuma Koizumi and Heiga Zen and Shigeki Karita and Yifan Ding and Kohei Yatabe and Nobuyuki Morioka and Michiel Bacchiani and Yu Zhang and Wei Han and Ankur Bapna},
  year      = {2023},
  booktitle = {INTERSPEECH 2023},
  pages     = {5496--5500},
  doi       = {10.21437/Interspeech.2023-1584},
  issn      = {2958-1796},
}

@inproceedings{zen2019libritts,
  title     = {{LibriTTS: A Corpus Derived from LibriSpeech for Text-to-Speech}},
  author    = {Heiga Zen and Viet Dang and Rob Clark and Yu Zhang and Ron J. Weiss and Ye Jia and Zhifeng Chen and Yonghui Wu},
  year      = {2019},
  booktitle = {Interspeech 2019},
  pages     = {1526--1530},
  doi       = {10.21437/Interspeech.2019-2441},
  issn      = {2958-1796},
}

@article{morrison2023cross,
  title={Cross-domain neural pitch and periodicity estimation},
  author={Morrison, Max and Hsieh, Caedon and Pruyne, Nathan and Pardo, Bryan},
  journal={arXiv preprint arXiv:2301.12258},
  year={2023}
}

@article{rautenberg2025speech,
  title={{Speech Synthesis along Perceptual Voice Quality Dimensions}},
  author={Rautenberg, Frederik and Kuhlmann, Michael and Seebauer, Fritz and Wiechmann, Jana and Wagner, Petra and Haeb-Umbach, Reinhold},
  journal={arXiv preprint arXiv:2501.08791},
  year={2025}
}

@inproceedings{casanova2022yourtts,
  title={{YourTTS: Towards Zero-Shot Multi-Speaker TTS and Zero-Shot Voice Conversion for Everyone}},
  author={Casanova, Edresson and Weber, Julian and Shulby, Christopher D and Junior, Arnaldo Candido and G{\"o}lge, Eren and Ponti, Moacir A},
  booktitle={International Conference on Machine Learning},
  pages={2709--2720},
  year={2022},
  organization={PMLR}
}

@inproceedings{kim2021conditional,
  title={{Conditional Variational Autoencoder with Adversarial Learning for End-to-End Text-to-Speech}},
  author={Kim, Jaehyeon and Kong, Jungil and Son, Juhee},
  booktitle={International Conference on Machine Learning},
  pages={5530--5540},
  year={2021},
  organization={PMLR}
}

@article{abdal2021styleflow,
  title={{StyleFlow: Attribute-conditioned Exploration of StyleGAN-Generated Images using Conditional Continuous Normalizing Flows}},
  author={Abdal, Rameen and Zhu, Peihao and Mitra, Niloy J and Wonka, Peter},
  journal={ACM Transactions on Graphics (ToG)},
  volume={40},
  number={3},
  pages={1--21},
  year={2021},
  publisher={ACM New York, NY}
}

@article{kong2020hifi,
  title={{HiFi-GAN}: Generative adversarial networks for efficient and high fidelity speech synthesis},
  author={Kong, Jungil and Kim, Jaehyeon and Bae, Jaekyoung},
  journal={Advances in neural information processing systems},
  volume={33},
  pages={17022--17033},
  year={2020}
}

@article{anastassiou2024voiceshop,
  title={{VoiceShop: A Unified Speech-to-Speech Framework for Identity-Preserving Zero-Shot Voice Editing}},
  author={Anastassiou, Philip and Tang, Zhenyu and Peng, Kainan and Jia, Dongya and Li, Jiaxin and Tu, Ming and Wang, Yuping and Wang, Yuxuan and Ma, Mingbo},
  journal={arXiv preprint arXiv:2404.06674},
  year={2024}
}

@article{chen2018neural,
  title={{Neural Ordinary Differential Equations}},
  author={Chen, Ricky TQ and Rubanova, Yulia and Bettencourt, Jesse and Duvenaud, David K},
  journal={Advances in neural information processing systems},
  volume={31},
  year={2018}
}

@inproceedings{Cord-Landwehr_Boeddeker_Zorilă_Doddipatla_Haeb-Umbach_2023, 
    title={{Frame-Wise and Overlap-Robust Speaker Embeddings for Meeting Diarization}}, 
    DOI={10.1109/icassp49357.2023.10095370}, 
    booktitle={ICASSP 2023-2023 IEEE International Conference on Acoustics, Speech and Signal Processing (ICASSP)}, 
    publisher={IEEE}, 
    author={Cord-Landwehr, Tobias and Boeddeker, Christoph and Zorilă, Cătălin and Doddipatla, Rama and Haeb-Umbach, Reinhold}, 
    year={2023}
}

@inproceedings{grathwohl2018ffjord,
  title={{FFJORD: Free-Form Continuous Dynamics for Scalable Reversible Generative Models}},
  author={Grathwohl, Will and Chen, Ricky TQ and Bettencourt, Jesse and Sutskever, Ilya and Duvenaud, David},
  booktitle={International Conference on Learning Representations},
  year={2019}
}

@inproceedings{lameris-etal-2024-role,
    title = "{The Role of Creaky Voice in Turn Taking and the Perception of Speaker Stance: Experiments Using Controllable {TTS}}",
    author = "Lameris, Harm  and
      Szekely, Eva  and
      Gustafson, Joakim",
    booktitle = "Proceedings of the 2024 Joint International Conference on Computational Linguistics, Language Resources and Evaluation (LREC-COLING 2024)",
    year = "2024",
    publisher = "ELRA and ICCL",
    pages = "16058--16065",
}

@inproceedings{radford2023robust,
  title={Robust speech recognition via large-scale weak supervision},
  author={Radford, Alec and Kim, Jong Wook and Xu, Tao and Brockman, Greg and McLeavey, Christine and Sutskever, Ilya},
  booktitle={International conference on machine learning},
  pages={28492--28518},
  year={2023},
  organization={PMLR}
}

@article{ITU808,
 title={{ITU-T Rec. P.}808, {S}ubjective evaluation of speech quality with a crowdsourcing approach},
 organization={International Telecommunication Union},
 year={2018}
}

@inproceedings{li2024articulatory,
  title     = {{GTR-Voice: Articulatory Phonetics Informed Controllable Expressive Speech Synthesis}},
  author    = {Zehua Kcriss Li and Meiying Melissa Chen and Yi Zhong and Pinxin Liu and Zhiyao Duan},
  year      = {2024},
  booktitle = {Interspeech 2024},
  pages     = {1775--1779},
  doi       = {10.21437/Interspeech.2024-2216},
}

@inproceedings{lameris2023prosody,
  title={Prosody-controllable spontaneous {TTS} with neural {HMM}s},
  author={Lameris, Harm and Mehta, Shivam and Henter, Gustav Eje and Gustafson, Joakim and Sz{\'e}kely, {\'E}va},
  booktitle={ICASSP 2023-2023 IEEE International Conference on Acoustics, Speech and Signal Processing (ICASSP)},
  pages={1--5},
  year={2023},
  organization={IEEE}
}

@inproceedings{lameris2023neural,
  title={{N}eural speech synthesis with controllable creaky voice style},
  author={Lameris, Harm and Wlodarczak, Marcin and Gustafson, Joakim and Sz{\'e}kely, {\'E}va},
  booktitle={International Congress of Phonetic Sciences (ICPhS)},
  pages={3141--3145},
  year={2023}
}

@inproceedings{rezende2015variational,
  title={{Variational Inference with Normalizing Flows}},
  author={Rezende, Danilo and Mohamed, Shakir},
  booktitle={International Conference on Machine Learning},
  pages={1530--1538},
  year={2015},
  organization={PMLR}
}

@article{bauer1972,
author = {David F. Bauer},
title = {{Constructing Confidence Sets Using Rank Statistics}},
journal = {Journal of the American Statistical Association},
volume = {67},
number = {339},
pages = {687--690},
year = {1972},
publisher = {ASA Website},
doi = {10.1080/01621459.1972.10481279},
}

@article{kreiman2007and,
  title={When and why listeners disagree in voice quality assessment tasks},
  author={Kreiman, Jody and Gerratt, Bruce R and Ito, Mika},
  journal={The Journal of the Acoustical Society of America},
  volume={122},
  number={4},
  pages={2354--2364},
  year={2007},
  publisher={AIP Publishing}
}

@article{hinterleitner2011evaluation,
  title={An evaluation protocol for the subjective assessment of text-to-speech in audiobook reading tasks},
  author={Hinterleitner, Florian and Neitzel, Georgina and M{\"o}ller, Sebastian and Norrenbrock, Christoph},
  journal={Proc. Blizzard Challenge workshop},
  year={2011},
  publisher={Citeseer}
}

@inproceedings{vctk2019,
  title={{CSTR VCTK Corpus: English Multi-speaker Corpus for CSTR Voice Cloning Toolkit (version 0.92)}},
  author={Yamagishi, Junichi and Veaux, Christophe and MacDonald, Kirsten},
  year={2019}
}

@article{tan2024naturalspeech,
  title={{Naturalspeech: End-to-end text-to-speech synthesis with human-level quality}},
  author={Tan, Xu and Chen, Jiawei and Liu, Haohe and Cong, Jian and Zhang, Chen and Liu, Yanqing and Wang, Xi and Leng, Yichong and Yi, Yuanhao and He, Lei and others},
  journal={IEEE Transactions on Pattern Analysis and Machine Intelligence},
  year={2024},
  publisher={IEEE}
}

@article{triantafyllopoulos2023overview,
  title={An overview of affective speech synthesis and conversion in the deep learning era},
  author={Triantafyllopoulos, Andreas and Schuller, Bj{\"o}rn W and {\.I}ymen, G{\"o}k{\c{c}}e and Sezgin, Metin and He, Xiangheng and Yang, Zijiang and Tzirakis, Panagiotis and Liu, Shuo and Mertes, Silvan and Andr{\'e}, Elisabeth and others},
  journal={Proceedings of the IEEE},
  volume={111},
  number={10},
  pages={1355--1381},
  year={2023},
  publisher={IEEE}
}

\end{document}